\documentclass[useAMS,usenatbib,usegraphicx]{mn2e}
\title[Mixed Morphology Supernova Remnants]{Simulations of Mixed Morphology Supernova Remnants With Anisotropic Thermal Conduction}
\author[D. A. Tilley, D. S. Balsara and J. C. Howk]{David A. Tilley\thanks{E-mail: dtilley@nd.edu (DAT); dbalsara@nd.edu (DSB); jhowk@nd.edu (JCH)}, Dinshaw S. Balsara and J. Christopher Howk\\
Department of Physics, University of Notre Dame, Notre Dame, Indiana 46556 USA}

\begin{document}
\pagerange{\pageref{firstpage}--\pageref{lastpage}}
\maketitle
\label{firstpage}

\begin{abstract}
We explore the role of anisotropic thermal conduction on the evolution of supernova remnants through interstellar media with a range of densities via numerical simulations.  We find that a remnant expanding in a dense environment can produce centre-bright hard x-ray emission within 20 kyr, and centre-bright soft x-ray emission within 60 kyr of the supernova event.  In a more tenuous environment, the appearance of a centre-bright structure in hard x-rays is delayed until about 60 kyr.  The soft x-ray emission from such a remnant may not become centre bright during its observable lifetime.  This can explain the observations that show that mixed-morphology supernova remnants preferentially occur close to denser, molecular environments.  Remnants expanding into denser environments tend to be smaller, making it easier for thermal conduction to make larger changes in the temperatures of their hot gas bubbles.  We show that the lower temperatures make it very favorable to use high-stage ions as diagnostics of the hot gas bubbles in SNRs.  In particular, the distribution of O \textsc{viii} transitions from shell-bright at early epochs to centre-bright at later epochs in the evolution of an SNR expanding in a dense ISM when the physics of thermal conduction is included.
\end{abstract}

\section{Introduction}

Supernovae play an important role in driving the thermal evolution of the interstellar medium (ISM henceforth).  The interiors of supernova remnants (SNR henceforth) are filled with extremely hot plasma, with temperatures of $10^6-10^7\;\mathrm{K}$.  The classical picture of the expansion of a supernova remnant into the interstellar medium identifies four stages that are still germane today \citep[see][]{woltjer72}.  At very early times, the remnant is freely expanding into the undisturbed medium.  After the shock sweeps up more mass than was ejected by the supernova, the remnant expands adiabatically according to the \citet{sedov59} solution.  When the cooling time of the matter behind the shock becomes comparable to the dynamical time of the expansion, a cold, dense shell is formed.  Eventually, the expansion velocity of the shock becomes comparable to the turbulent velocity in the ISM, and the remnant disperses.

SNRs can be classified into several morphological types; see \citet{weiler_sramek88} and \citet{jones_etal98} for excellent reviews.  The classical or shell-type SNRs are the most common and are shell-bright in both x-rays and radio emission.  The former arises from thermal bremsstrahlung and inner-shell metal line emission in the hot gas behind the shock.   Synchrotron emission from cosmic rays and magnetic field swept up by the shock leads to the radio morphology.  Plerionic remnants are powered by a central pulsar and have bright, nonthermal radio emission in their interior.  Finally, there is a composite class, including radio and x-ray composite morphologies, which exhibit both a radio-bright shell and centre-bright x-ray emission.  The x-ray composite remnants (also known as thermally composite or mixed morphology, MM henceforth) comprise 8 per cent of all remnants \citep{rho_petre98}.  

Rho \& Petre also found that MM SNRs preferentially occur toward the Galactic midplane where the gas is denser.  Rho \& Petre cite \citet{frail_etal96,reach_rho96,koo_moon97} to demonstrate that the shocks from MM SNRs could be interacting with molecular cloud material, suggesting that MM SNRs occur in denser environments.  In this paper, we examine that hypothesis.  \citet{kawasaki_etal05} have proposed all SNRs pass through a composite phase. Their argument is based on over-ionisation in the majority of MM SNRs. We examine the hypothesis that all SNRs pass through an MM phase with our simulations.

Two existing models for the formation of MM SNRs have figured prominently since the recognition of this class of remnant.  The first model, proposed by \citet{white_long91}, has cool, dense clouds that become engulfed by the supernova shock and hot gas bubble.  Thermal conduction and turbulent mixing transfer heat between the bubble and clouds, cooling the interior of the SNR and increasing its density as the cold gas in the clouds evaporates and mixes with the remnant.  

The second model by \citet{cox_etal99} proposed that thermal conduction acting by itself can cause a SNR to become centre-bright in x-rays if the remnant is expanding into a dense medium.  As the sound speed of the thermal electrons is significantly greater than the sound speed of the ions, the electrons can propagate faster than the outer shock and carry energy away from the interior.  The interior density increases as the bubble cools \citep{slavin_cox92,cox_etal99,shelton_etal99}.  The transport of heat from the hot interior of the remnant to the surrounding medium leads to lower temperatures and higher densities inside the SNR, leading to greater emission in x-rays from the centre of the remnant.  

In each of these models, thermal conduction is invoked to transport energy from the hot gas to cool gas.  They differ primarily in how they distribute the cold, dense gas component.  In either of these scenarios, the higher densities and cooler temperatures of $10^5-10^6$ K in the hot gas bubble favor the production of soft x-rays in the centres of these remnants.  We concentrate on the latter model in this work.

In the presence of a magnetic field, thermal conduction becomes anisotropic as the free electrons that are responsible for the bulk of the heat transport will not propagate further than the Larmor radius of the magnetic field \citep{spitzer_pfig56,cowie_mckee77a,balbus86}.  Thus, the geometry of the magnetic field cannot be neglected when calculating the thermal heat flux in a multidimensional calculation.  As a result, \citet{tilley_balsara06} examined the role of anisotropic thermal conduction in the presence of magnetic fields and showed that it is important in determining the filling factor of hot gas in our Galaxy.

\citet{slavin_cox92,slavin_cox93} used a simplified representation of the magnetic field in their early calculations and we improve on that here.  \citet{velazquez_martinell_raga_giacani04} include the anisotropy in the thermal conduction introduced by the magnetic field, but do not include the Lorentz force of the magnetic field on the gas. As a consequence, the magnetic field was stirred by the turbulent motions developing behind the supernova shock, greatly inhibiting thermal conduction. Without the feedback of the magnetic field on the gas, however, this can only be regarded as a lower limit on the degree of thermal conduction.   We removed this restriction in \citet{tilley_balsara06}, and here we examine further consequences of removing that restriction.

A goal of our work, with its limited but interesting set of input micro-physics, is to produce diagnostics that may be compared with observations. Recent work by the \textit{FUSE}, \textit{Chandra} and \textit{XMM} observatories have provided insight into the total amount of hot gas in the Galaxy, as traced by highly-ionised stages of oxygen, esp. O \textsc{vi} and O \textsc{viii} \citep{oegerle_etal05, yao_wang05,savage_lehner06}. These ions probe gas in the temperature range $\sim10^5$ K to $\sim10^8$ K, providing valuable diagnostics of the thermal state of the hot material in the Milky Way.  While such data have so far been used to test global models of the ISM \citep{mckee_ostriker77, korpi_etal99, kim_etal01,avillez_breitschwerdt04, balsara_etal04, balsara_kim05, avillez_breitschwerdt05,maclow_etal05,tilley_balsara06}, it is also possible to use high-stage ions as diagnostics for the structure of SNRs.  Indeed, while earlier observatories allowed spectroscopy of small regions of supernova remnants \citep[e.g.,][]{vedder_etal86}, recent \textit{Chandra} and \textit{XMM} observations have provided spatially-resolved measurements of O \textsc{vii} and O \textsc{viii} emission from SNRs \citep{gaetz_etal00, rasmussen_etal01, vanderheyden_etal02, flanagan_etal04}. We further develop the idea of using high-stage ions as diagnostics of hot gas in SNRs in this paper.

In this paper we report on the results of a series of MHD simulations on the evolution of SNRs with anisotropic thermal conduction.  We explore a parameter space of initial ISM conditions in density, temperature, and magnetic field strength (Section \ref{section_numerical}).  In Section \ref{section_mmr} we analyse the effects of the environment on the evolution of the remnant, in particular focusing upon the effects of the environment on producing mixed-morphology remnants.  Section \ref{section_ions} explores the role that high-stage ions could play in diagnosing the hot gas.  Section \ref{section_conclusion} provides some conclusions.

\section{Numerical Setup}\label{section_numerical}
\begin{table}
\caption{Initial conditions of the ISM in our simulations. $\rho$ is the density in amu $\mathrm{cm}^{-3}$; T is the temperature in Kelvin; B is the magnetic field in $\mu$G; and L is the grid size in parsecs.\label{table_ic}}
\begin{tabular}{lcccc}
Run & $\rho$ & T & B & L\\
\hline 
L0 & 0.7 & 8000 & 0.0 & 75\\
L1 & 0.7 & 8000 & 3.0 & 75\\
M0 & 1.0 & 10000 & 0.0 & 50\\
M1 & 1.0 & 10000 & 3.0 & 50\\
H0 & 5.0 & 10000 & 0.0 & 50\\
H1 & 5.0 & 10000 & 6.0 & 50\\ \hline
\end{tabular}
\end{table}
\begin{figure*}
\includegraphics[width=42mm]{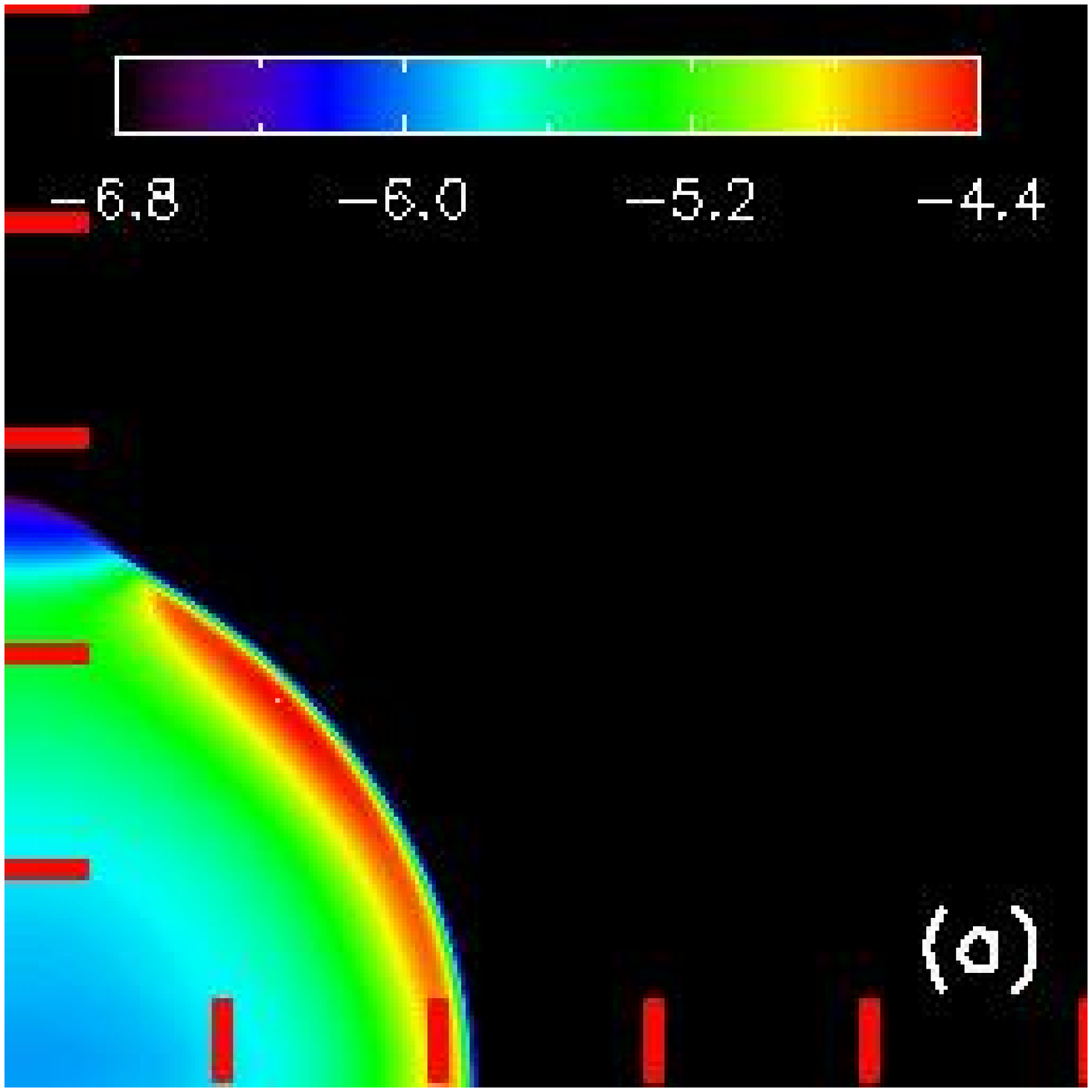}\includegraphics[width=42mm]{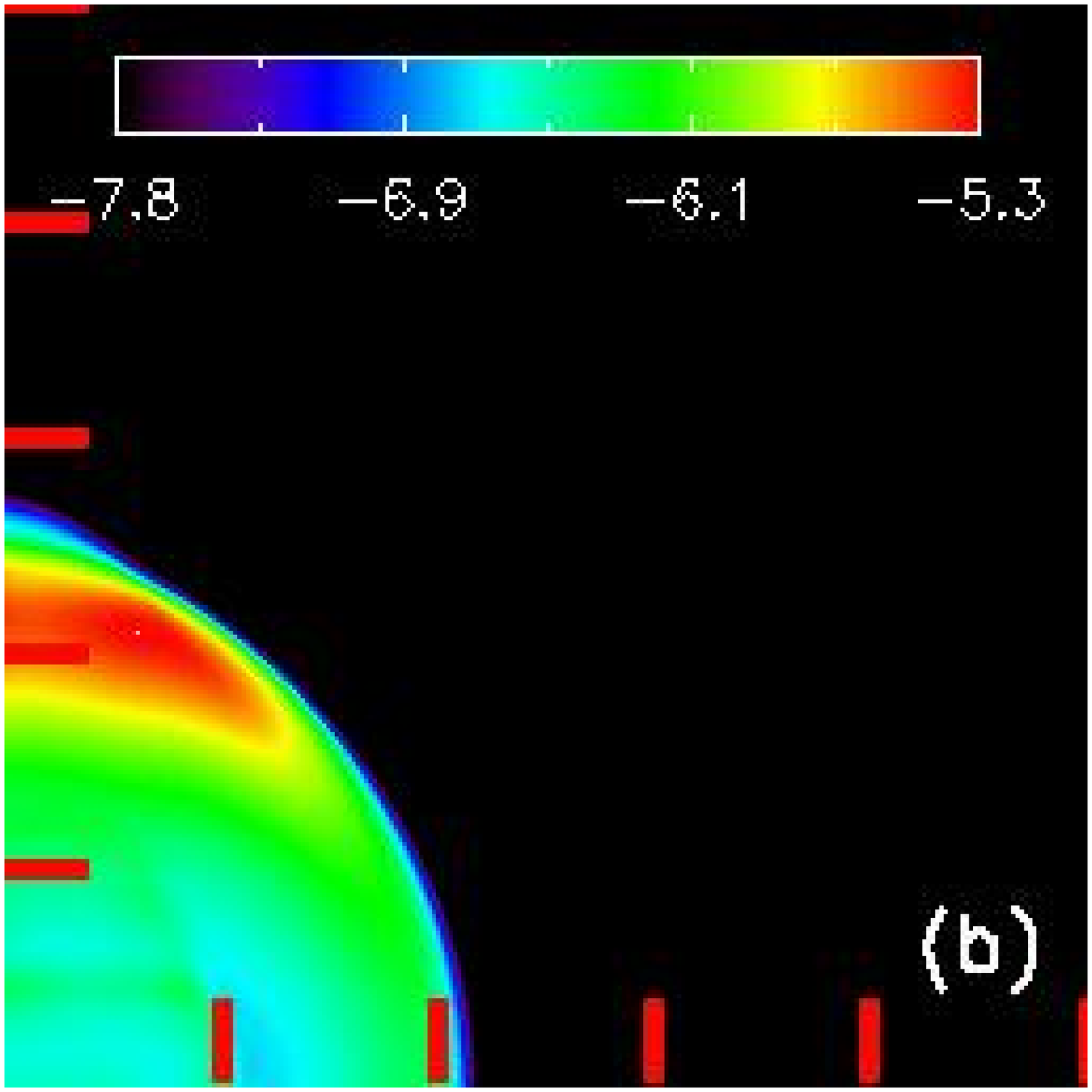}\includegraphics[width=42mm]{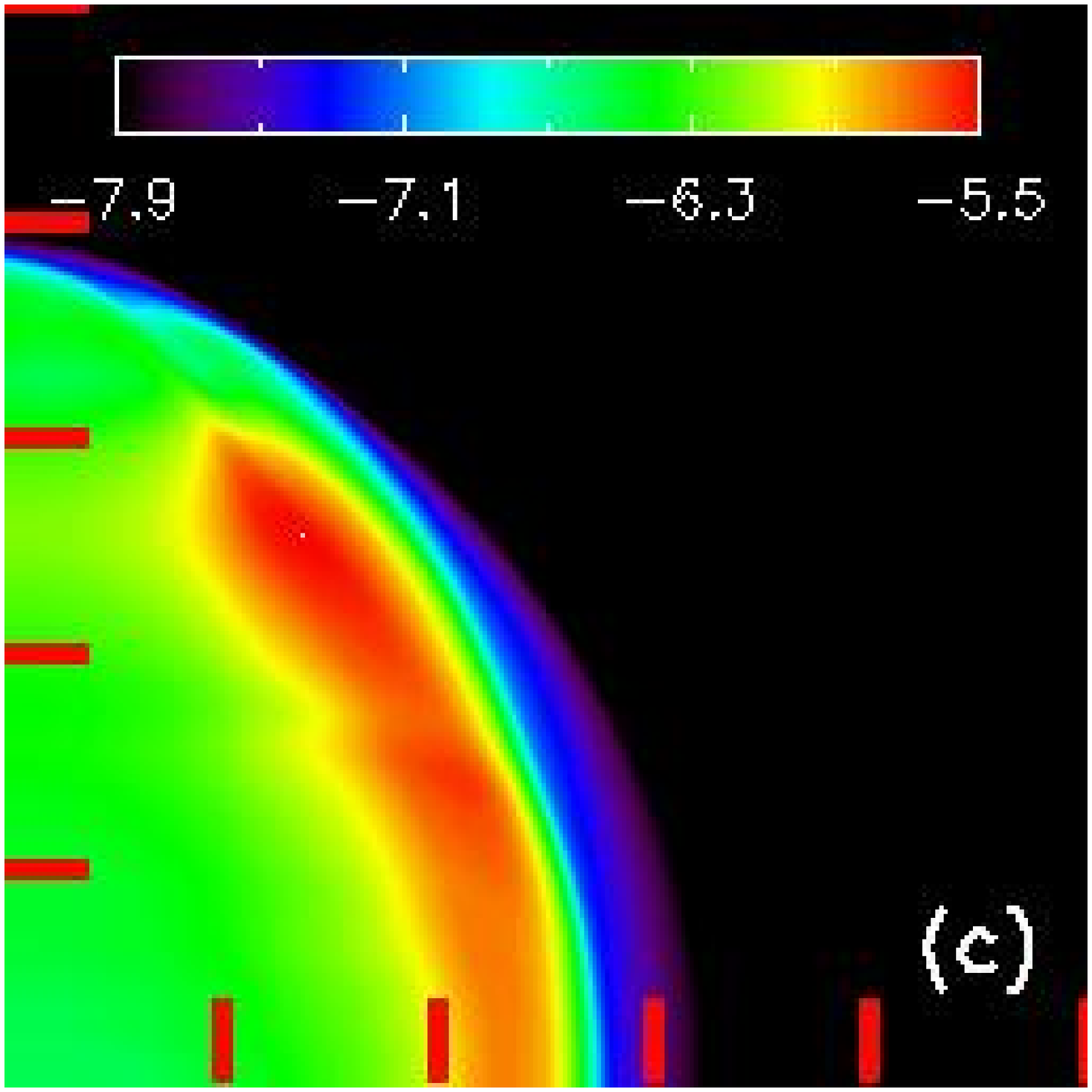}\includegraphics[width=42mm]{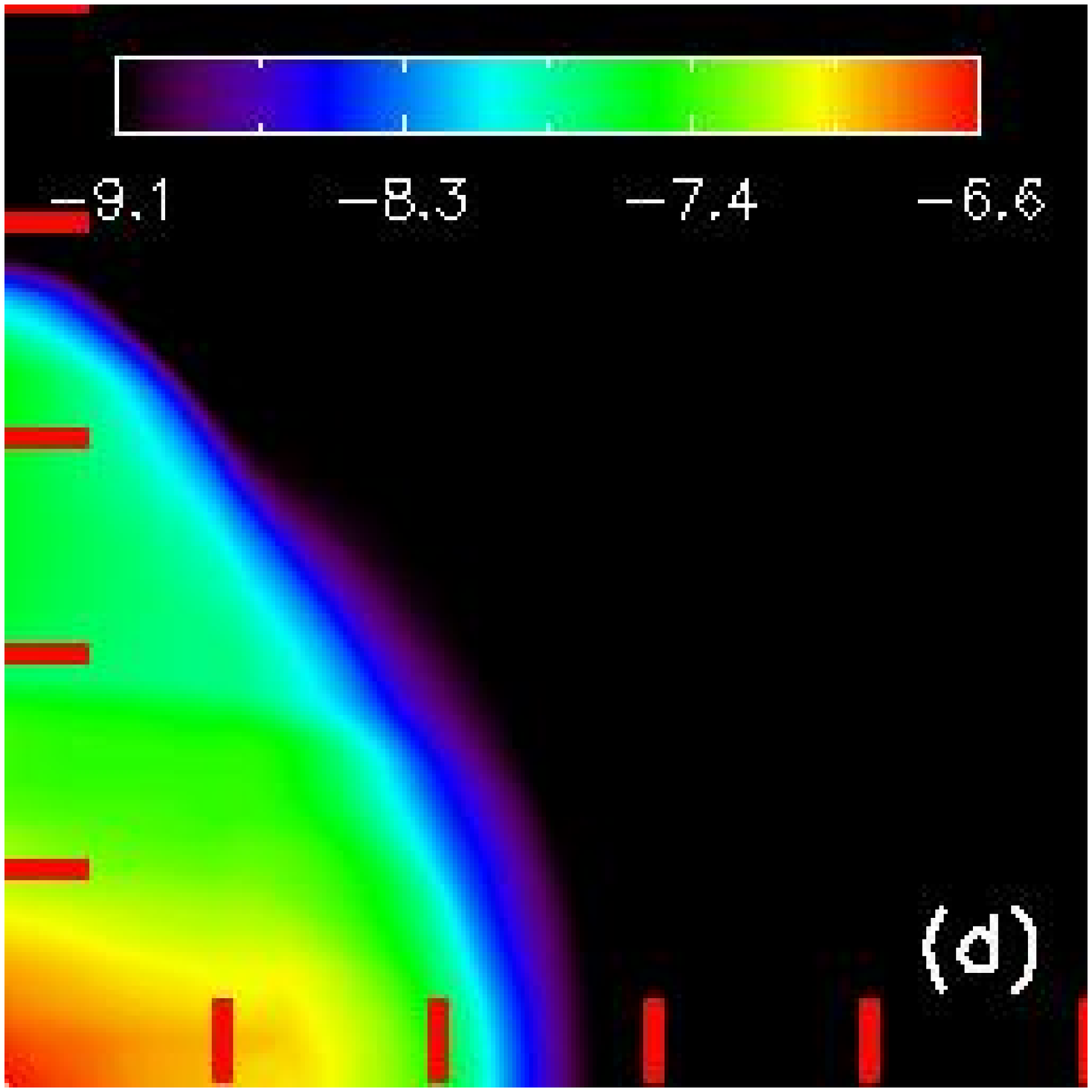}\\
\includegraphics[width=42mm]{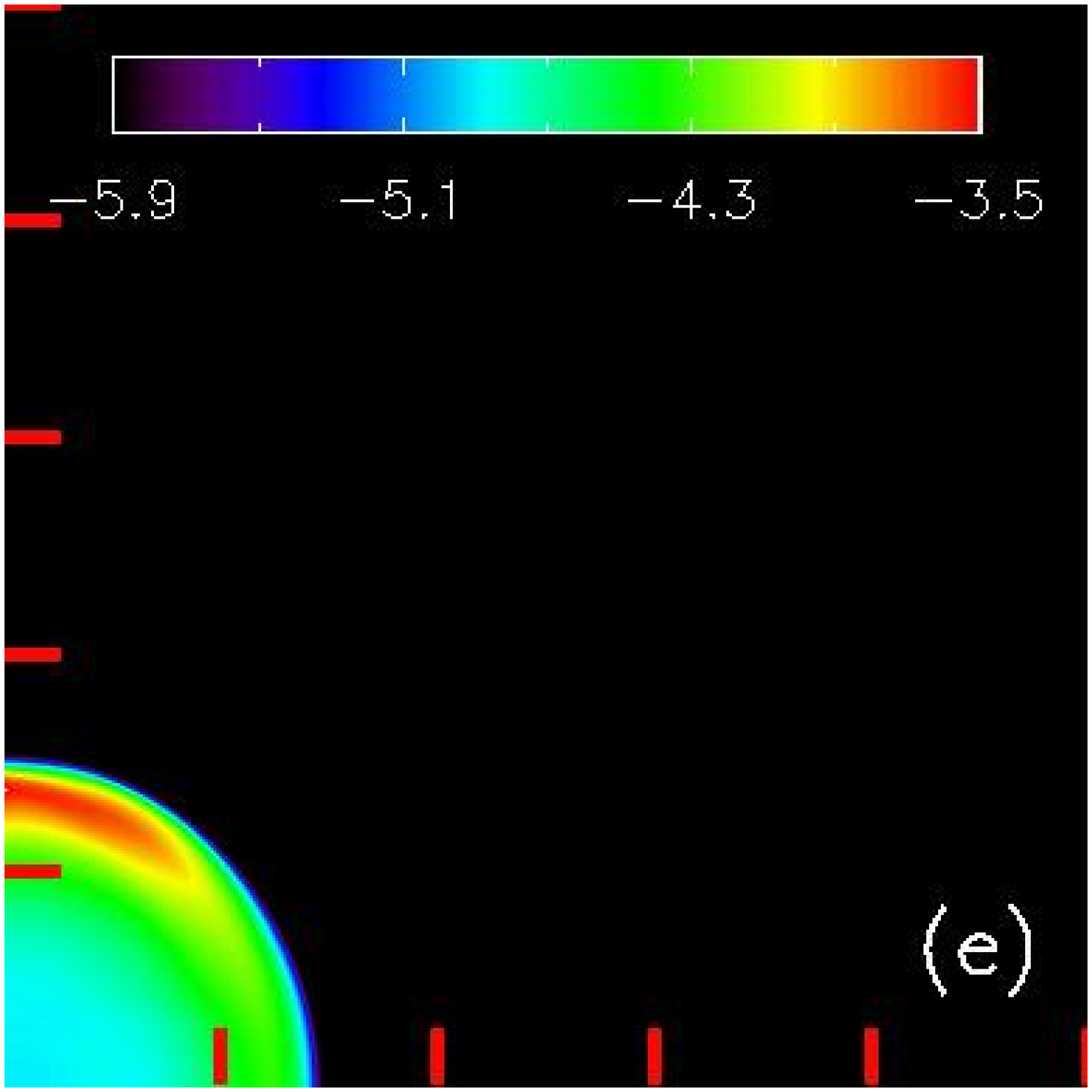}\includegraphics[width=42mm]{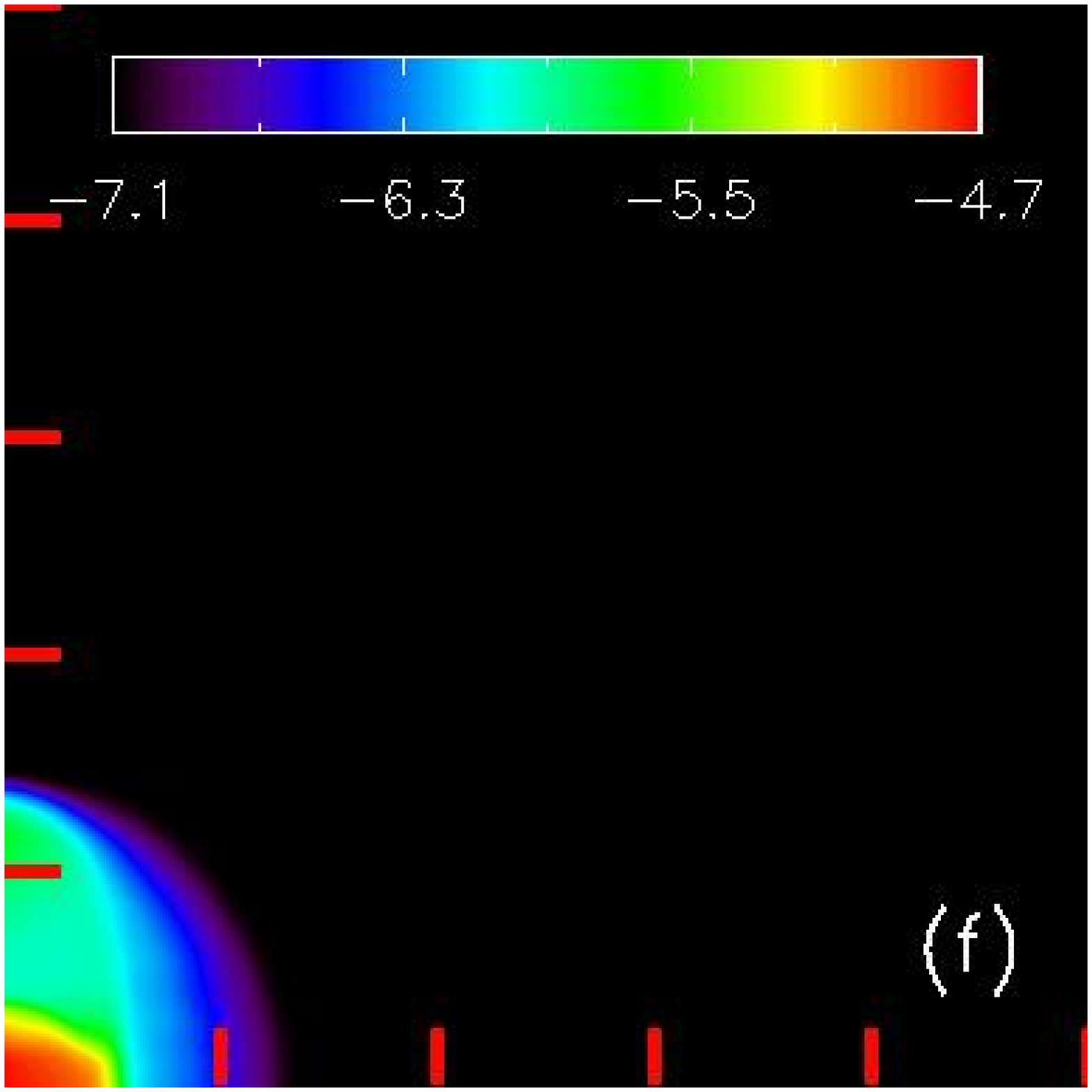}\includegraphics[width=42mm]{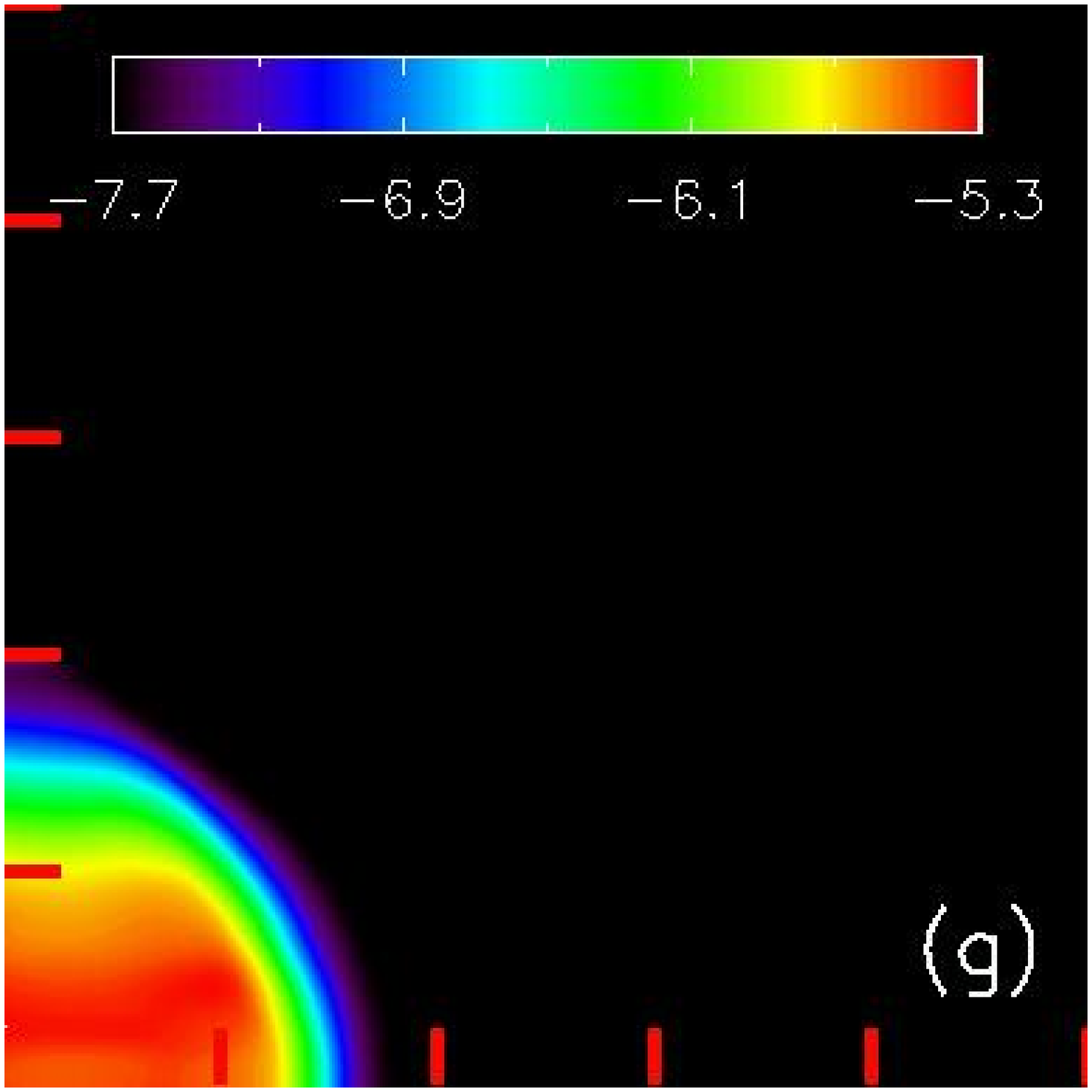}\includegraphics[width=42mm]{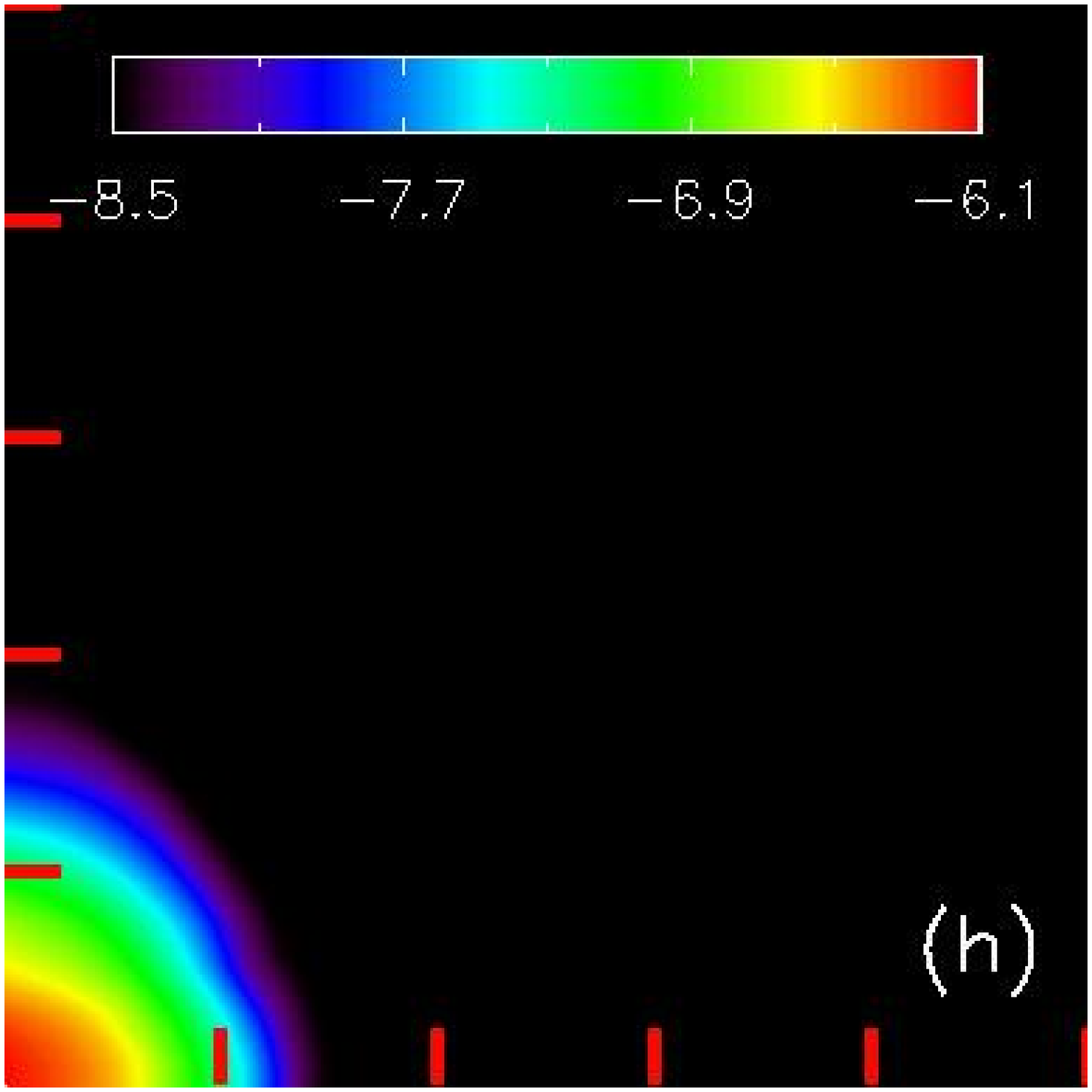}
\caption{Simulated x-ray surface brightness in soft and hard x-rays, as described in the text.  (a) and (b) show soft and hard x-rays from Run L1 (low-density ISM) at 20 kyr.  (c) and (d) do the same for Run L1 at 60 kyr.  (e) and (f) show soft and hard x-rays from Run H1 (high-density ISM) at 20 kyr, while (g) and (h) show the same for Run H1 at 60 kyr. The tick marks indicate projected distances of 10 pc.  The colour scale is the logarithm of the surface brightness in ergs s$^{-1}$ cm$^{-2}$ sr$^{-1}$.\label{fig_mmr}}
\end{figure*}

We carried out our simulations by using the \textsc{riemann} code \citep{roe_balsara96,balsara98a,balsara98b,balsara_spicer99a,balsara_spicer99b,balsara01a,balsara04}. \textsc{riemann} updates the MHD equations using a second-order accurate dimensionally unsplit TVD scheme.  We use an implicit solver for the thermal conduction.  We discuss the details of our treatment of anisotropic thermal conduction in an accompanying paper (Balsara et al., in preparation).  We draw on \citet{spitzer_pfig56,cowie_mckee77a,balbus86,slavin_cox92} for our description of the classical thermal conduction flux, which is proportional to a temperature-dependent conductivity and the projection of the thermal gradient on to the local magnetic field:
\begin{eqnarray}
\mathbf{F}_\mathrm{clas} & = & - a T^{5/2} \mathbf{\hat{b}}\left(\mathbf{\hat{b}}\cdot\nabla T\right)\label{eq_flux_classical}
\end{eqnarray}
where T is the temperature, $a=6\times10^{-7}\;\mathrm{g\;cm}^{-3}\;\mathrm{s}^{-3}\;\mathrm{K}^{-7/2}$ and $\mathbf{\hat{b}}=\mathbf{B}/|\mathbf{B}|$ is a unit vector along the magnetic field.  When the flux becomes saturated, it no longer depends on the value of the temperature gradient, only on the presence of a parallel component of the temperature gradient to the magnetic field \citep{balbus86}:
\begin{eqnarray}
\mathbf{F}_\mathrm{sat} & = & -5\phi\rho c_s^3 \mathrm{sgn}\left(\mathbf{\hat{b}}\cdot\nabla T\right)\mathbf{\hat{b}}\label{eq_flux_saturated}
\end{eqnarray}
where $c_s$ is the sound speed, $\rho$ the density and $\phi\sim0.3$ \citep{balbus_mckee82}.  A flux limiter is utilized to transition from the classical flux to the saturated flux \citep{balbus_mckee82,slavin_cox92}.  The transition from the classical (Eq. \ref{eq_flux_classical}) to saturated (Eq. \ref{eq_flux_saturated}) thermal conduction causes the flux to change character from parabolic to hyperbolic \citep{balbus86}.  The changing character in the flux of thermal conduction is accounted for by using a Newton-Krylov multigrid method that correctly incorporates the anisotropic structure of the thermal conduction operator in the presence of a magnetic field.  We have tested our formulation of the thermal conduction against the supernova shock radius-time relation of \citet{cioffi_mckee_bertschinger88} and found excellent agreement.  We also found very good agreement with the instability growth rates predicted by \citet{field65}.  These tests are described in detail in (Balsara et al., in preparation).

Radiative cooling is calculated using the cooling function of \citet{macdonald_bailey81}.  We balance the cooling rate of our initial conditions with a constant heating rate, such that the initial conditions would remain at constant temperature in the absence of the supernova remnant.  Our choice of equilibrium cooling permits us to do a large number of runs without carrying the large number of species that would be necessary in a formulation that retains the non-equilibrium ionisation \citep{slavin_cox92}.  However, more economical formulations for the inclusion of non-equilibrium cooling are becoming available \citep{benjamin_benson_cox01} and it is our intent to incorporate such physics in future models.  The focus of our present paper is to examine the role of thermal conduction, which does not depend sensitively on the details of the radiative cooling.

We use a uniform mesh in cylindrical coordinates of 384x384 zones in the R-z plane to update the MHD equations.  We chose our initial densities and temperatures of the quiescent environment so that they are representative of interesting interstellar medium conditions.  Our lowest density runs, labelled ``L0'' and ``L1'', have a density of 0.7 amu cm$^{-3}$, a temperature of 8000 K and a magnetic field of 0 $\mu$G and 3 $\mu$G, respectively.  Our intermediate runs, labelled ``M0'' and ``M1'', have a density of 1.0 amu cm$^{-3}$, a temperature of 10000 K and a magnetic of 0 $\mu$G and 3 $\mu$G, respectively.  Our highest density runs, labelled ``H0'' and ``H1'', have a density of 5.0 amu cm$^{-3}$, a temperature of 10000 K and a magnetic field of 0 $\mu$G and 6 $\mu$G, respectively.  These initial conditions are summarized in Table \ref{table_ic}.  In all of these cases, we initialize our supernova with $10^{51}$ ergs of energy, partitioned such that one-third of the supernova energy is contained in thermal energy, and the other two-thirds in kinetic energy.  We initialize our remnant with a radius of 10 zones in all of our simulations.  In order to follow the complete evolution of the remnant, we wish to ensure that the outer shock remains on the grid at all times.  The physical size of our grid in each of the simulations in Table \ref{table_ic} reflects this.  To serve as a counterpoint we have carried out a complementary set of simulations without thermal conduction, but including radiative cooling.

\section{Mixed Morphology Remnants}\label{section_mmr}
\begin{figure}
\includegraphics[width=42mm]{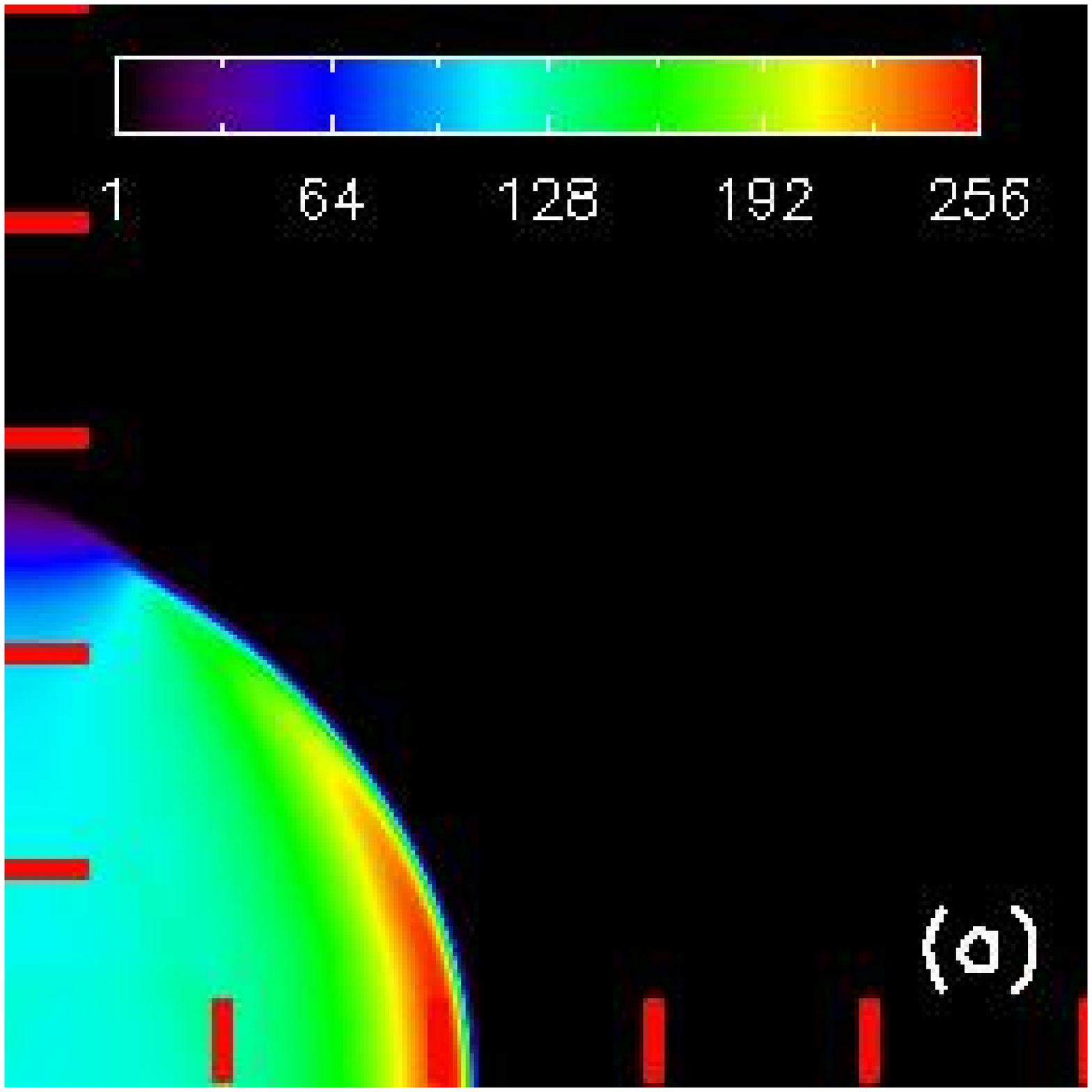}\includegraphics[width=42mm]{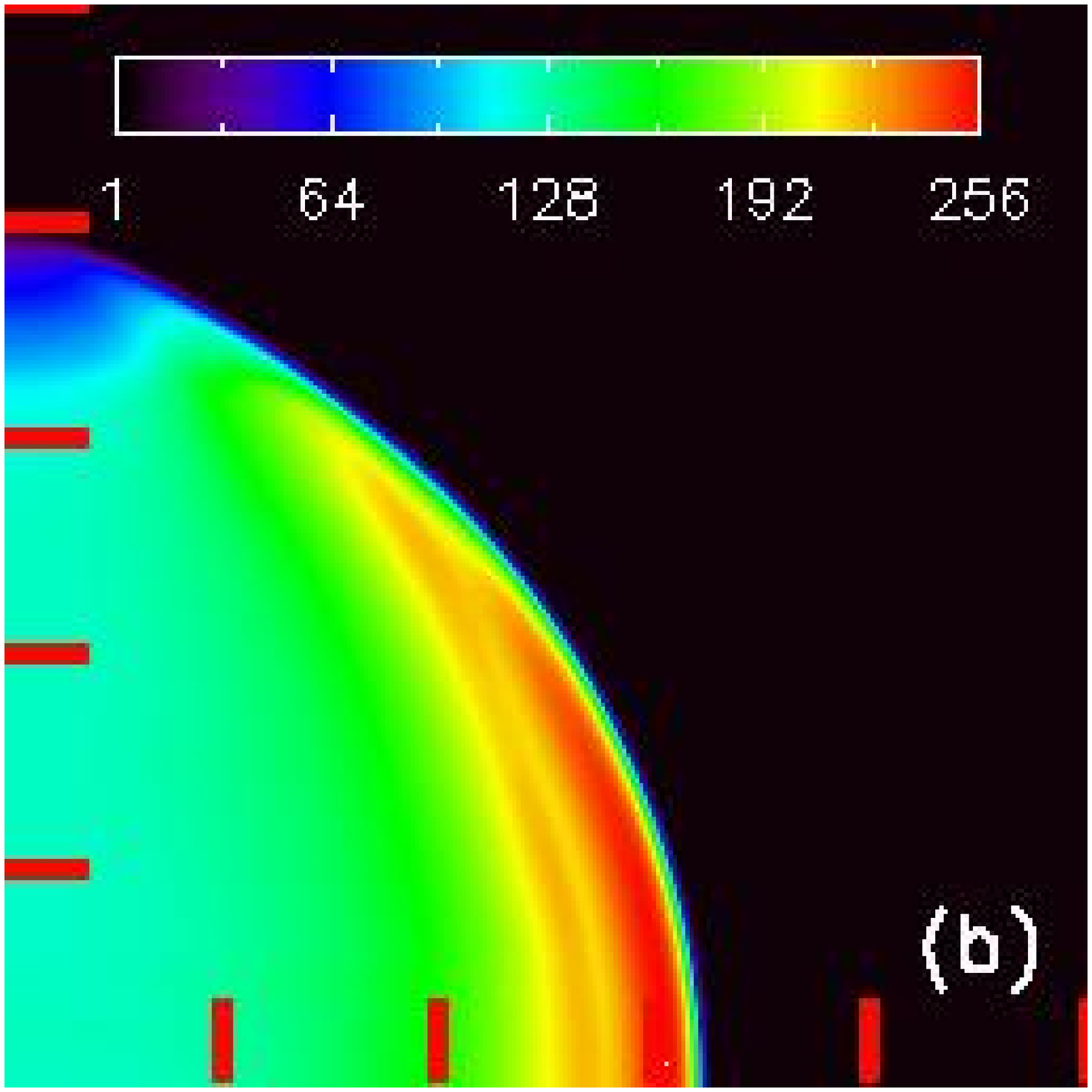}\\
\includegraphics[width=42mm]{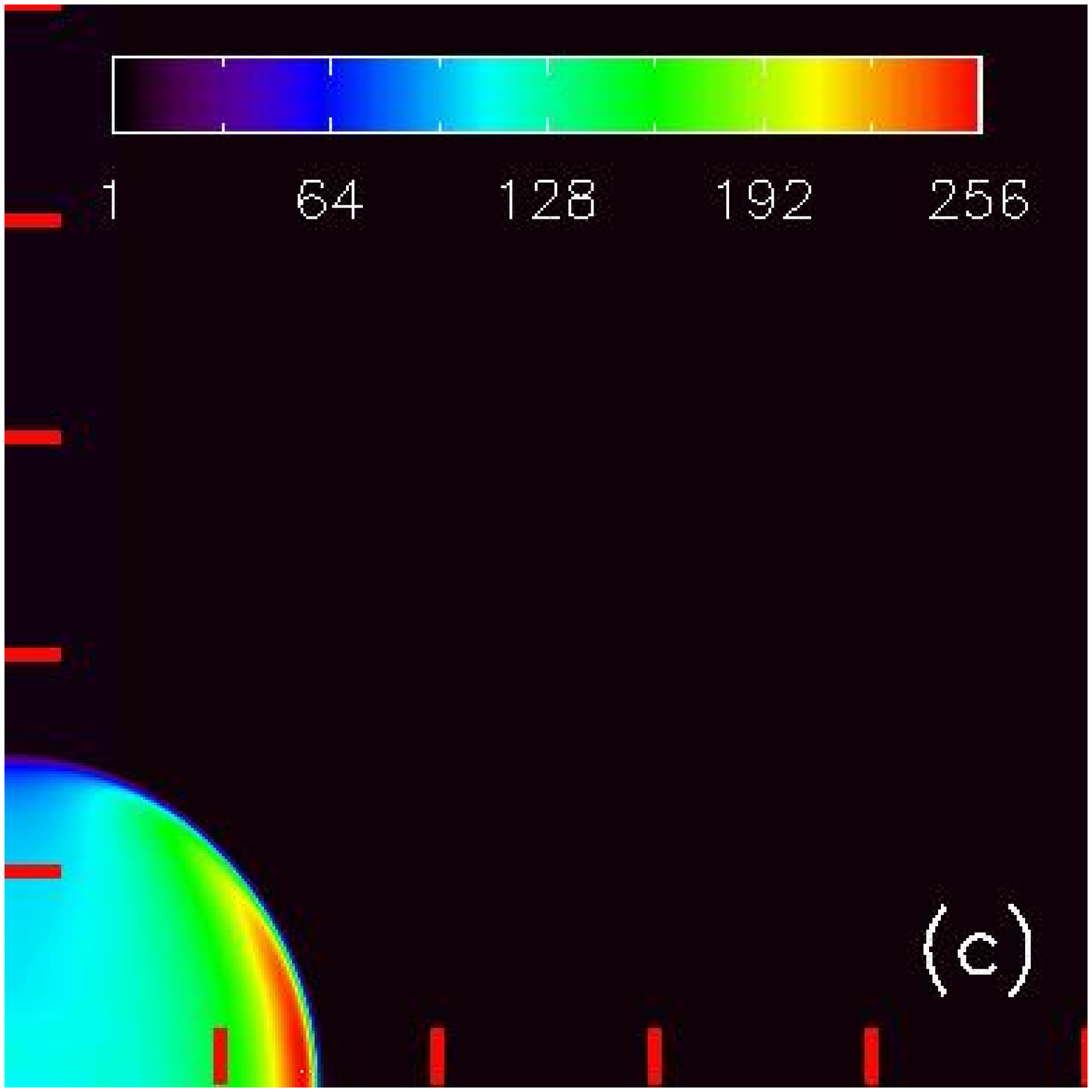}\includegraphics[width=42mm]{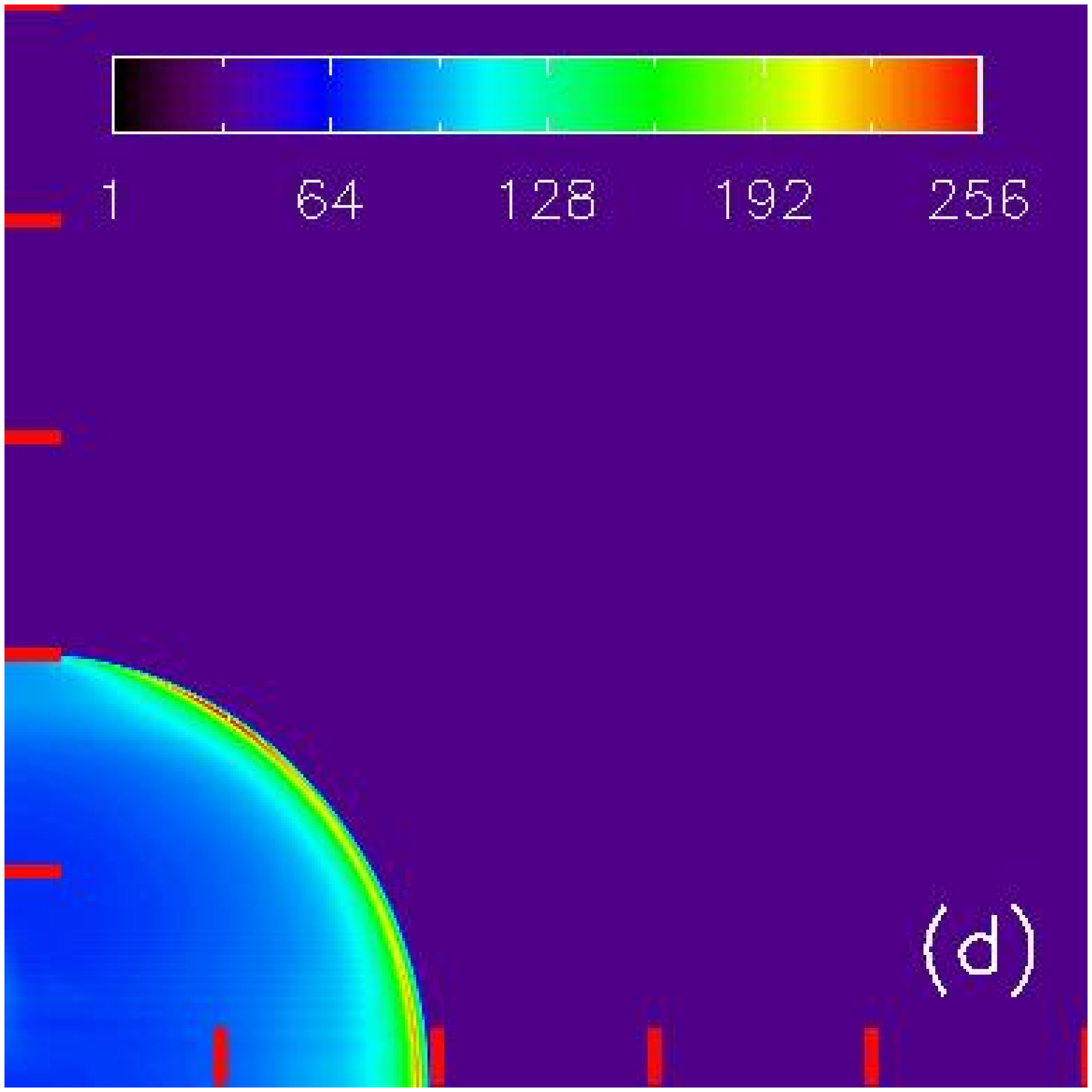}
\caption{Simulated synchrotron surface brightness $\propto PB^{1.5}$ from SNRs, shown at two epochs, (Left) 20 kyr after the explosion.  (Right) 60 kyr after the explosion.  The top row illustrates the expansion of a SNR into a standard ISM (Run L1).  The bottom row illustrates the expansion of a SNR into a higher-density ISM (Run H1).  The tick marks indicate projected distances of 10 pc.  Units on the linear colour scale are arbitrary, and scaled such that the greatest surface brightness has a value of 256.\label{fig_mmr_radio}}
\end{figure}
\begin{figure}
\includegraphics[width=84mm]{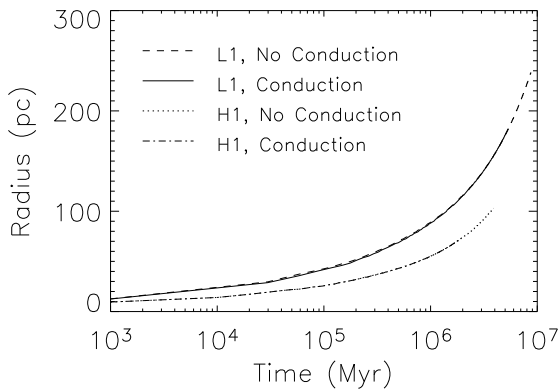}
\caption{Radius of the outer shock as it evolves with time, showing the difference in the evolution between runs with and without conduction, and runs in low or high density ISMs.  The solid and dashed lines virtually overlie each other.  Likewise, the dotted and dot-dashed lines also practically overlie each other.\label{fig_outershock}}
\end{figure}

The structure of supernova remnants that include anisotropic thermal conduction in magnetized ISMs has been studied in \citet{tilley_balsara06}.  In their fig. 1, they intercompare remnants that include thermal conduction to those that exclude thermal conduction, but in both cases they incorporate the effects of radiative cooling.  In the snowplow phase, they find that the temperature of the hottest gas is reduced by a factor of $\sim 10$ , while the density is increased by the same factor, in those remnants that include thermal conduction.  They also find that the inclusion of a uniform magnetic field restricts the thermal conduction in the direction of the magnetic field.  As a consequence, the structure of the hot gas bubble becomes elongated in the direction of the magnetic field.

MM SNRs are primarily diagnosed in x-ray and radio emission. Moreover, \textit{Chandra} has made it possible to image the same remnant in soft and hard x-rays with high spatial resolution. To examine the influence of thermal conduction on the production of the morphology of the remnant at these energies, we produce simulated emission maps of our models at several epochs during the course of the SNR's evolution.  We calculate the x-ray emissivity for  $10^5-10^8$ K gas using \citet{raymond_smith77}.  These emissivities include the contribution from bremsstrahlung, high-order atomic lines, recombination and two-photon emission for photons in the range of 100 eV - 10 keV.  We then integrated the optically thin emission along the line-of-sight to create projected intensity maps of the simulated SNRs in x-rays.

The projected soft and hard x-ray maps are shown in Fig. \ref{fig_mmr}.  To retain some concordance with \textit{Chandra} data, we define 'soft x-rays' as those with photon energies in the range of 300-800 eV, and 'hard x-rays' as those with energies in the range of 1-5 keV. The resultant images were not convolved with an instrumental spatial or spectral response function. Fig. \ref{fig_mmr} (a) and (b) show the soft and hard x-rays in the low-density run L1 at 20 kyr.  Fig. \ref{fig_mmr} (c) and (d) do the same at 60 kyr.  Fig. \ref{fig_mmr} (e) through (h) show similar data for the high-density run H1 at the same epochs.  Fig. \ref{fig_mmr} (a) and (b) show that at early epochs, the low-density run L1 is shell-bright in soft x-rays, and also shell-bright in hard x-rays.  The magnetic field draws heat from the hottest central gas and carries it to the pole, making Fig. \ref{fig_mmr}(b) brighter at the pole, while Fig. \ref{fig_mmr}(a) is brighter at the equator.  Fig. \ref{fig_mmr} (e) and (f) show that at 20 kyr the high-density run H1 is shell-bright in the soft x-rays, but is already centre-bright in the hard x-rays.  \citet{shelton_etal99} also present simulated hard x-ray images of remnants expanding into an unmagnetized medium of similar density, finding that they too produce centre-bright images in the hard x-ray and thereby producing a concordance between our data and theirs.

At later epochs, i.e. 60 kyr, Fig. \ref{fig_mmr} (c) and (d) show that the low-density run L1 is still shell-bright in soft x-rays, but is becoming centre-bright in hard x-rays.  Fig. \ref{fig_mmr} (g) and (h) show that the high-density run H1 has become centre-bright in both soft and hard x-rays.    This shows that thermal conduction plays an important role in making model H1 centre-bright in x-rays.  A similar conclusion emerges when we cross-compare models L0 and H0 without magnetic fields.  The difference between models that include magnetic fields and those that exclude it is that without magnetic fields thermal conduction operates in both the R and z directions, thereby increasing the rate at which energy is transported from the hot gas bubble.  As a result, thermally conducting remnants that exclude magnetic fields make their transition to a centre-bright x-ray morphology at an earlier epoch.

The morphological differences between Fig. \ref{fig_mmr} (a) and (b) require further explanation.  The anisotropy in the thermal conduction flux caused by the presence of the magnetic field leads to the incomplete appearance of the shell.  Temperature variations perpendicular to the field are not reduced due to thermal conduction.  
As the magnetic field is swept up by the SNR, the magnetic flux becomes concentrated in a thin shell.  Most of the magnetic field in the interior of the remnant nevertheless retains some memory of its original orientation.  As a result, it is preferentially attached to the exterior medium closer to the z-axis.  Consequently, the outer regions of the hot bubble near the axis of symmetry have a lower density and higher temperature than the outer parts of the bubble near the midplane, resulting in lower emission in soft x-rays and a higher emission in hard x-rays near the pole at early epochs.  A remnant going off into a denser medium is smaller and reaches its snowplow phase faster.  As a result, thermal conduction is more efficient at transferring energy within such a remnant.  For that reason, we see that H1 has already become centre-bright in hard x-rays, with the soft x-rays showing an incomplete shell, see Fig. \ref{fig_mmr}(e) and (f).  We see, therefore, that the choice of wavebands used to image the remnant can also make a difference in its classification as a centre-bright remnant in x-rays.  In particular, remnants expanding into a denser ISM would be much more likely to be classified as centre-bright if they are imaged in hard x-rays instead of soft x-rays.

Magnetic fields are compressed by the outer shock in an SNR, and can even be amplified if the shock interacts with interstellar turbulence \citep{balsara_benjamin_cox01}.  Cosmic rays can also be swept up by the outer shock and are most likely accelerated by it.  We do not include cosmic ray acceleration processes here.  As a result, we calculate synchrotron emissivities as $\epsilon_\mathrm{synch}\propto P B^{1.5}$ \citep[see][]{clarke_burns_norman89,jun_jones99}.  The simulated synchrotron maps are illustrated in Fig. \ref{fig_mmr_radio}.  In both the high-density and low-density environments the concentrated magnetic field appears as a bright shell behind the shock at all times.  Thus, we find a clear difference between the morphological appearance at 60 kyr in x-rays and radio emission.

The previous paragraph has shown us that the radio morphology is always shell-bright.  We have also seen that the transition to a centre-bright x-ray morphology depends to some extent on the wavebands used.  If we use soft x-rays, we find that remnants expanding into a denser environment enter a mixed-morphology phase within about 60 kyr, while remnants expanding into a standard ISM do not transition to a mixed morphology within that time.  However, even the remnants expanding into a standard ISM do transition to a mixed morphology after about 90 kyr.  This might, however, be too late in the life of a remnant for it to be visible.  We note that the soft x-ray flux has already decreased by an order of magnitude between Fig. \ref{fig_mmr}(a) and Fig. \ref{fig_mmr}(c).  We expect it to decrease by a similar magnitude by 90 kyr.  If we use hard x-rays, we find that the remnant expanding into a denser environment is already centre-bright by about 20 kyr, while the remnant expanding into a standard ISM does not become centre-bright until about 60 kyr.  We see, therefore, that the classification as a mixed-morphology remnant might depend strongly on the waveband used.  We can, nevertheless, say with confidence that remnants expanding into denser ISMs might spend a significant fraction of their observable lives as mixed-morphology remnants.  We cannot say the same for remnants expanding into a standard ISM.  As a result, we conclude that the conjecture in \citet{kawasaki_etal05}, that all remnants pass through a mixed-morphology phase in their evolution, may in theory be quite plausible.  However, remnants expanding into ISMs with standard densities or below-average densities may not display themselves as mixed-morphology remnants during a significant phase of their observable lifetimes.

It is also important to realize that thermal conduction does not reduce the total energy, it only transports it from one place to another. Consequently, our numerical formulation is also fully conservative. Fig. \ref{fig_outershock} shows the location of the outer shock along the R-axis as a function of time. There is clearly no difference in the size of the shock whether thermal conduction is included or not.  Since the location of the outer shock (at least in the radial direction) is unaffected by thermal conduction, we can conclude that the post-shock pressure should be the same in runs with and without conduction. The interior of the hot gas bubble is always hot enough (and the sound speed in it high enough) to ensure that the bubble achieves pressure balance with the post-shock gas. In the presence of thermal conduction the hot gas bubble can transport part of its thermal energy to some of the swept-up post-shock gas.  This process can, therefore, lower the temperature in the hot gas bubble. To maintain constancy of pressure, the density in the hot gas will be correspondingly higher.

\section{High-Stage Ions as Diagnostics of the Hot Gas}\label{section_ions}
\begin{figure}
\includegraphics[width=84mm]{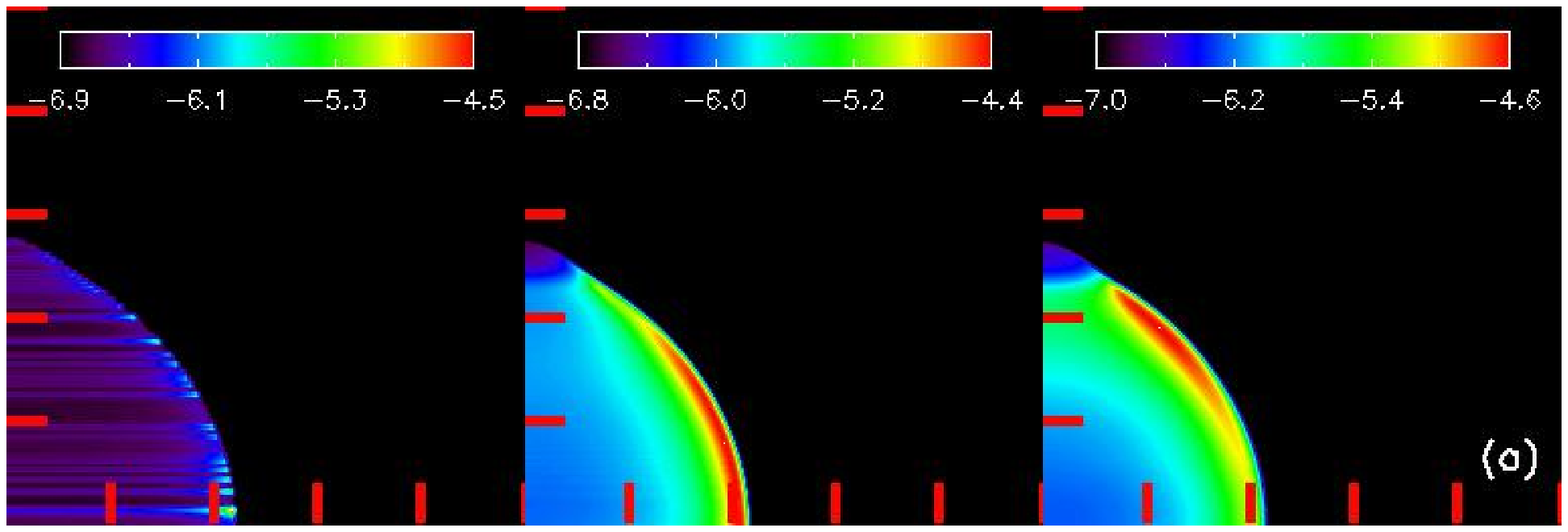}\\
\includegraphics[width=84mm]{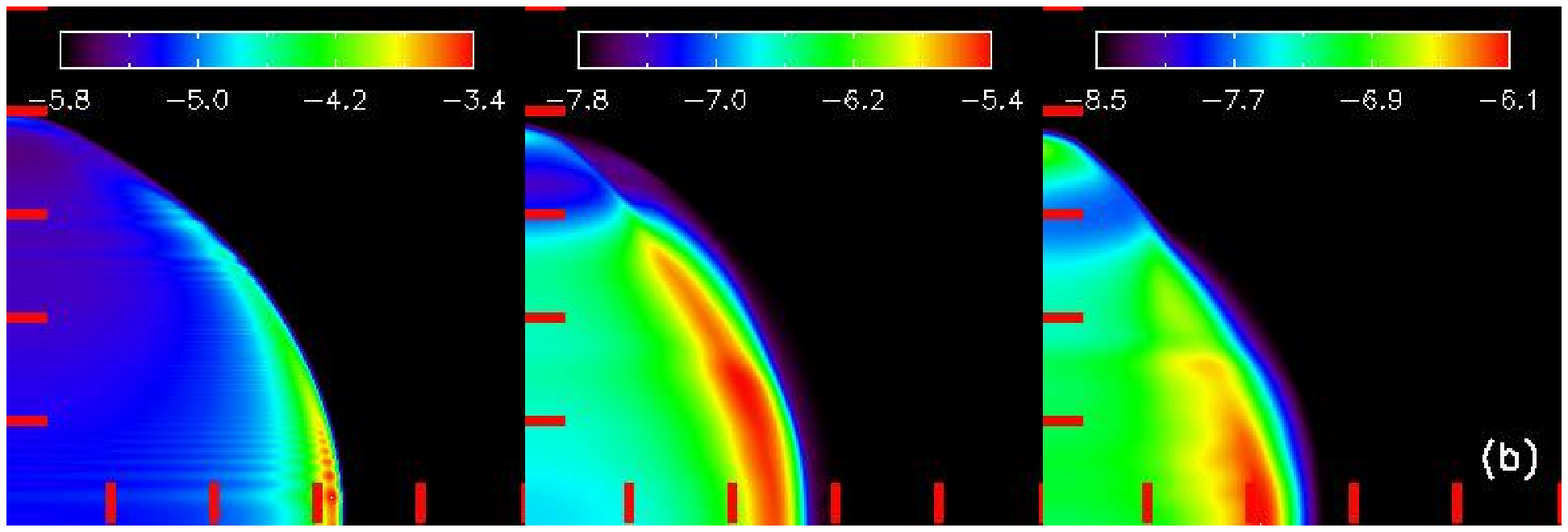}\\
\includegraphics[width=84mm]{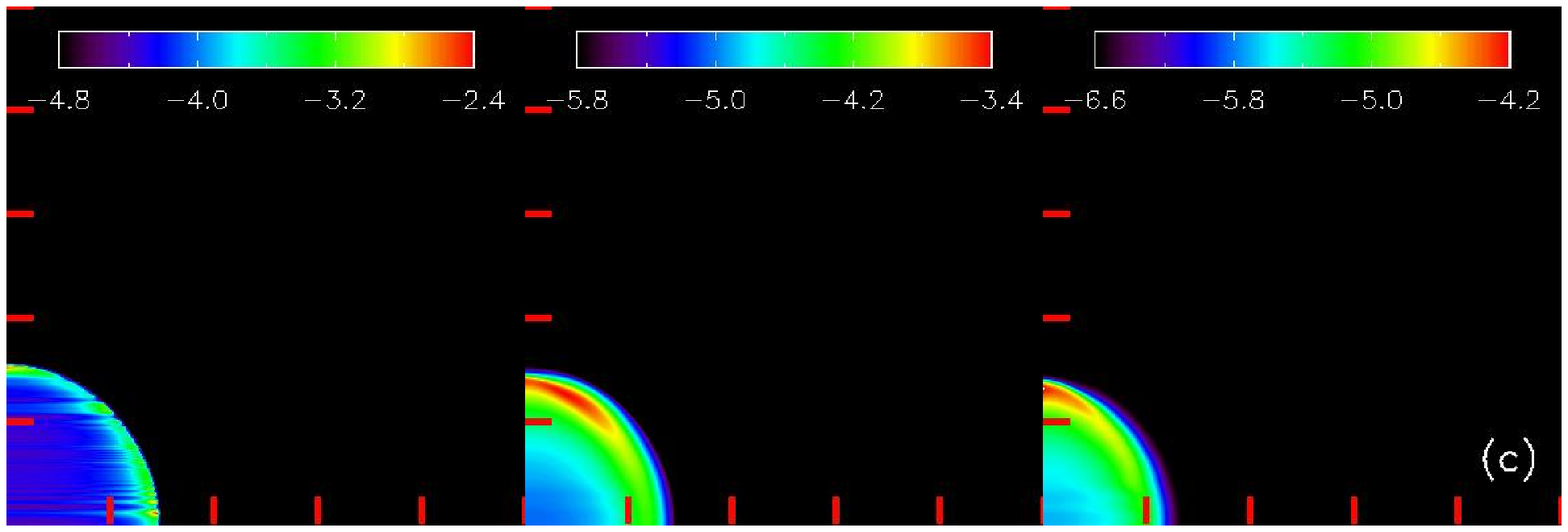}\\
\includegraphics[width=84mm]{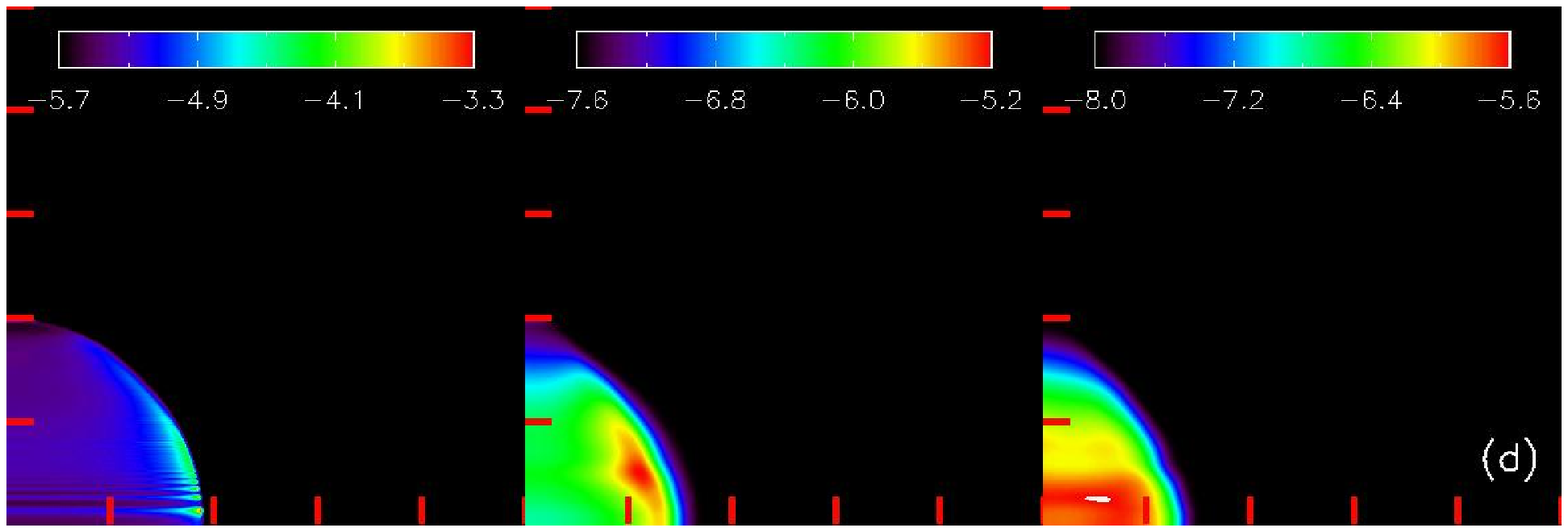}
\caption{Surface brightness due to emission from oxygen ions for (a) Run L1 at 20 kyr, (b) Run L1 at 60 kyr, (c) Run H1 at 20 kyr, and  (d) Run H1 at 60 kyr.  The ions are (left) O \textsc{vi}, (centre) O \textsc{vii}, and (right) O \textsc{viii}. The tick marks indicate projected distances of 10 pc.  Because the distribution of O \textsc{vi} emitting gas originates from a very narrow shell, the image shows striations that arise from the pixellation of this boundary in the simulations.   The colour scale is the logarithm of the surface brightness in ergs s$^{-1}$ cm$^{-2}$ sr$^{-1}$. \label{fig_oxygen}}
\end{figure}
\begin{figure}
\includegraphics[width=84mm]{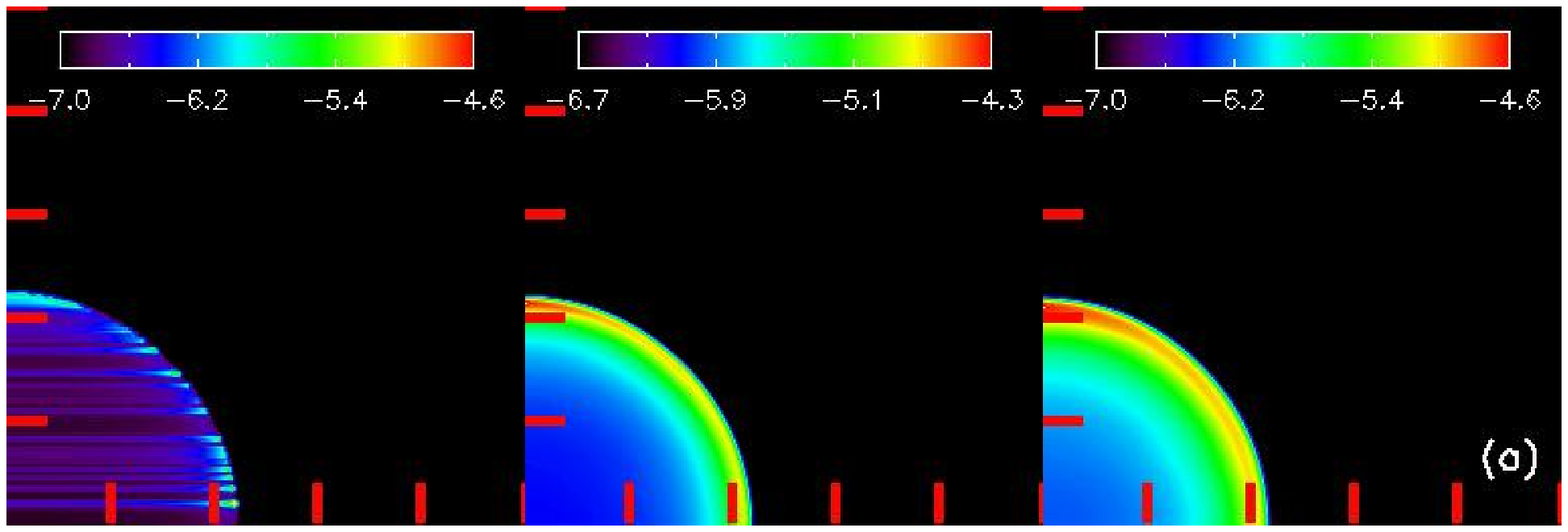}\\
\includegraphics[width=84mm]{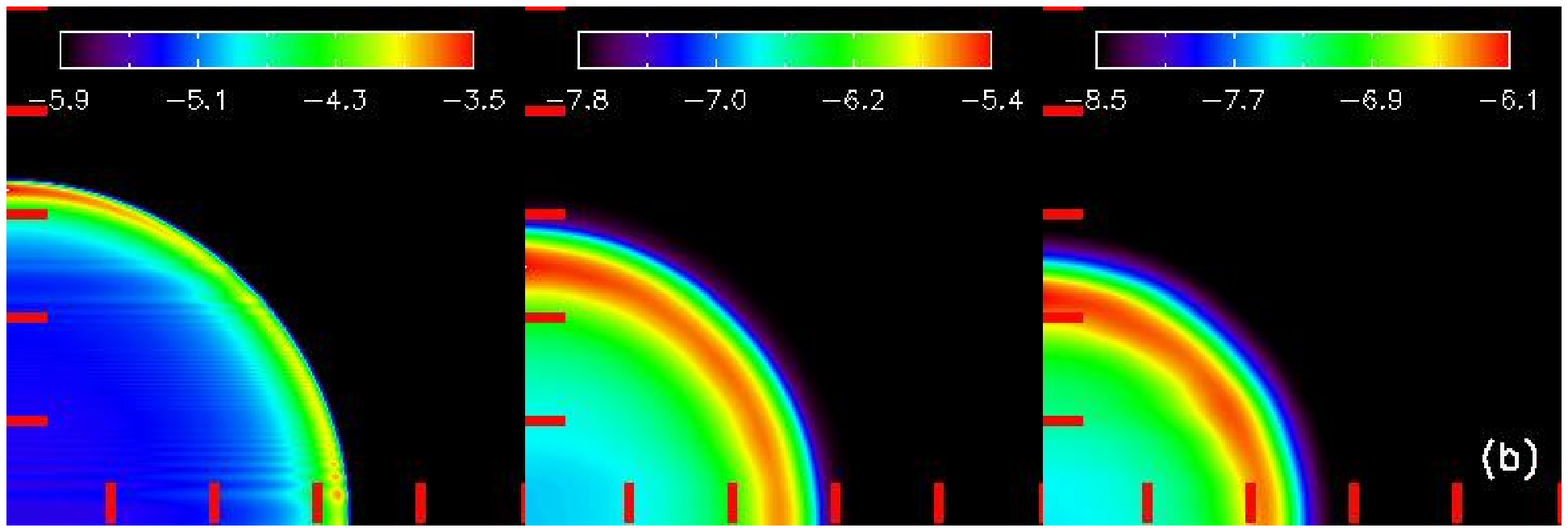}\\
\includegraphics[width=84mm]{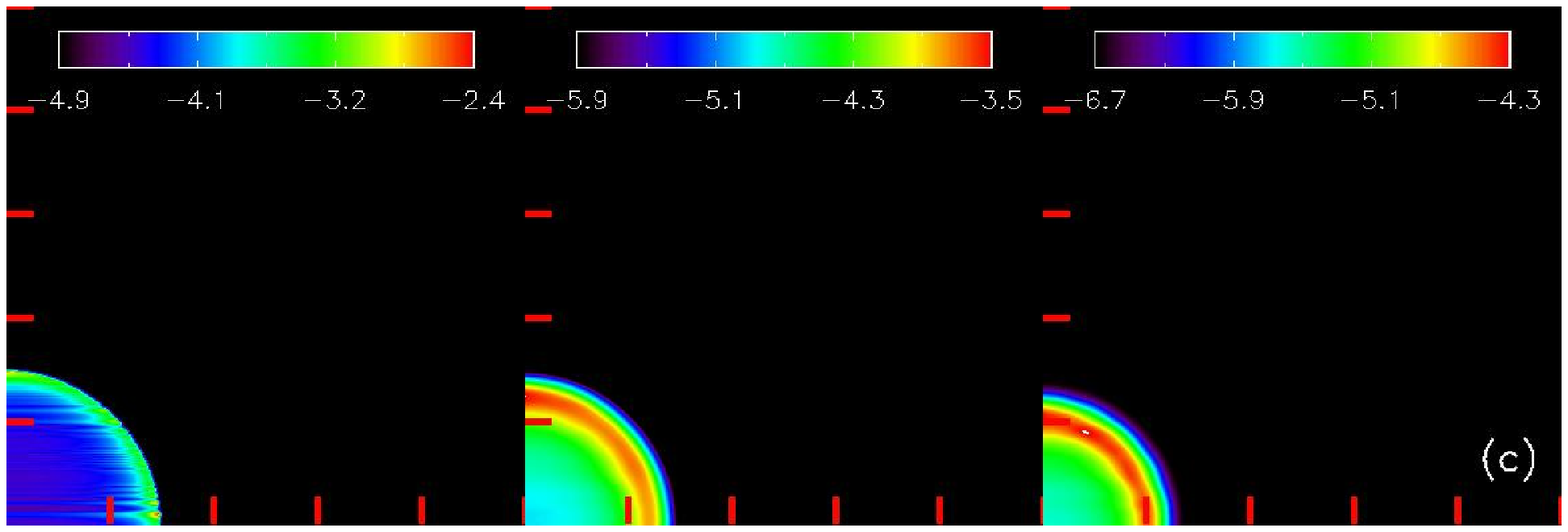}\\
\includegraphics[width=84mm]{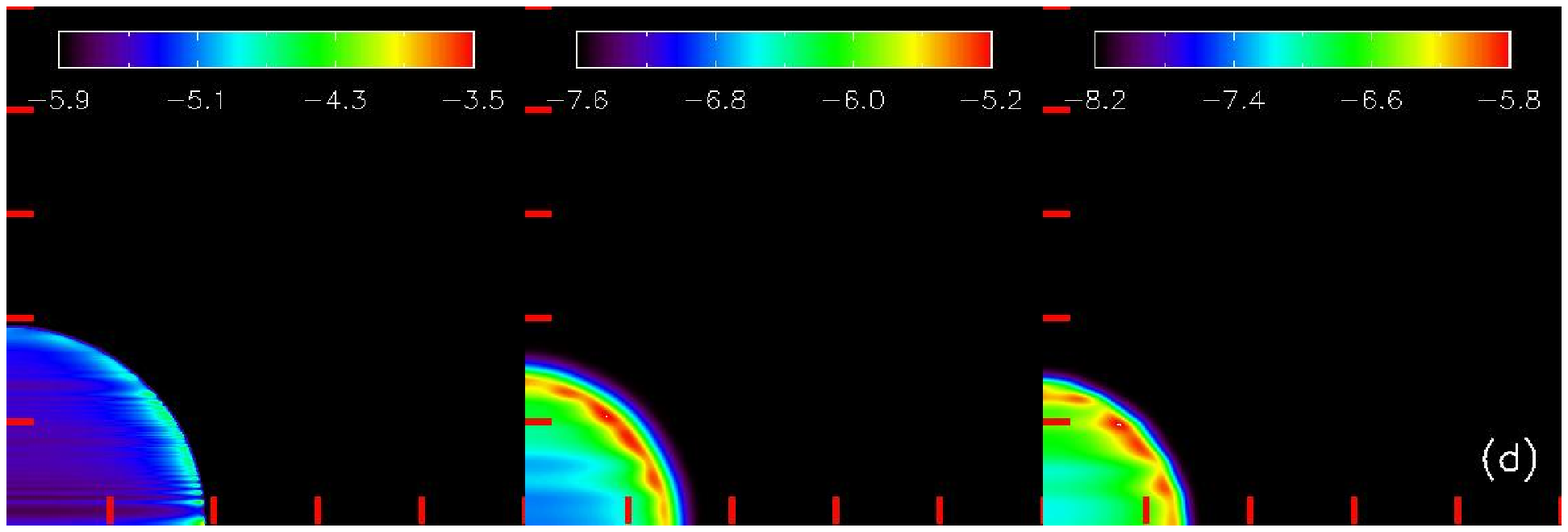}
\caption{Surface brighness due to emission from oxygen ions for simulations without thermal conduction.  (a) Run L1 at 20 kyr, (b) Run L1 at 60 kyr, (c) Run H1 at 20 kyr, and  (d) Run H1 at 60 kyr.  The ions are (left) O \textsc{vi}, (centre) O \textsc{vii}, and (right) O \textsc{viii}.  The tick marks indicate projected distances of 10 pc.  Because the distribution of O \textsc{vi} emitting gas originates from a very narrow shell, the image shows striations that arise from the pixellation of this boundary in the simulations.  The colour scale is the logarithm of the surface brightness in ergs s$^{-1}$ cm$^{-2}$ sr$^{-1}$. \label{fig_oxygen_nc}}
\end{figure}

In this Section, we examine the role of high-stage ions as diagnostics of hot gas in SNRs.  Highly-ionised oxygen is  readily observed in X-rays (e.g., through the O \textsc{vii} ``triplet'' at 21.60, 21.80, and 22.09 \AA\ and O \textsc{viii} Lyman-$\alpha$ at 18.98 \AA) and in the far-ultraviolet (using the O \textsc{vi} doublet at 1033.8 \AA).  We will calculate the emission from these ions which probe a range of temperatures (O \textsc{vi} peaks at $\sim 3\times 10^5$ K, O \textsc{vii} at $ 2.8\times 10^5$ -- $1.8\times 10^6$ K, and O \textsc{viii} at $\sim 2.2 \times 10^6$ K). Thus, these ions provide good diagnostics of gas in these temperature regimes, which are important during the remnant evolution.

We use the line emissivity calculations from the Astrophysical Plasma Emission Code \citep[APEC;][]{smith_etal01} with atomic data from the Astrophysical Plasma Emission Database (APED) included in the ATOMDB distribution (v. 1.3.1).\footnote{Available via   http://cxc.harvard.edu/atomdb/.}  These calculations assume collisional ionisation equilibrium \citep{mazzotta_etal98}, and we adopt an abundance of log O/H$=-3.34$ from the solar system \citep{asplund_etal04}.  In what follows, O \textsc{vi} emission represents the sum of the O \textsc{vi} doublet emission, while O \textsc{vii} refers to the summed emission from the aforementioned triplet; O~\textsc{viii} emission refers to Lyman-$\alpha$ 18.98 \AA\ emission.  We have tested the temperature-dependent line emissivities from the APEC models using a treatment of O \textsc{vi} emission following \citet{shull_slavin94}, and we find very good agreement.

Fig. \ref{fig_oxygen} and \ref{fig_oxygen_nc} show maps of O \textsc{vi}, O \textsc{vii}, and O \textsc{viii} at various epochs in simulations of SNRs going through  low and high density ISMs.  The results shown in Fig. \ref{fig_oxygen} include thermal conduction.  The results in Fig. \ref{fig_oxygen_nc} do not include thermal conduction.  In each instance we show simulated images of O \textsc{vi}, O \textsc{vii}, and O \textsc{viii} from left to right.  Fig. \ref{fig_oxygen}(a) and Fig. \ref{fig_oxygen_nc}(a) show these data in the low-density run at 20 kyr, and Fig. \ref{fig_oxygen}(b) and Fig. \ref{fig_oxygen_nc}(b) show the same simulation at 60 kyr.  Similarly, Fig. \ref{fig_oxygen}(c) and Fig. \ref{fig_oxygen_nc}(c) show these data in the high-density run at 20 kyr, and Fig. \ref{fig_oxygen}(d) and Fig. \ref{fig_oxygen_nc}(d) show the same simulation at 60 kyr.

We see that O \textsc{vi} is always concentrated at the boundary of the hot gas bubble, thereby showing its utility as a tracer of the gas that it transitional between the hot and cold phases, and as a tracer of turbulent mixing layers as shown by \citet{borkowski_balbus_fristrom90,begelman_mckee90}.  The O \textsc{vii} shows itself to be shell-bright in all instances, with and without thermal conduction, indicating that it is not a good tracer of the role of thermal conduction in SNRs.  Physically, this may be due to the broad range of temperatures over which O \textsc{vii} is the dominant ionisation stage of oxygen. By contrast, O \textsc{viii} probes a narrow range of higher temperatures, and the distribution of O \textsc{viii} is most interesting. 
Fig. \ref{fig_oxygen}(d) shows that the O \textsc{viii} is centrally brightened at later stages in the evolution of the SNR when thermal conduction is included.  This trend occurs to a lesser extent in Fig. \ref{fig_oxygen}(b), with more O \textsc{viii} emission arising in the central region when thermal conduction is included.  Fig. \ref{fig_oxygen_nc}(b) and (d) show a strongly edge-brightened appearance in O \textsc{viii}, even at late epochs, when thermal conduction is not included.  Thus, thermal conduction plays an important role in cooling down the interior of the hot gas bubble at late epochs.  This changes the ionisation balance, driving the interior from O \textsc{ix} to O \textsc{viii} and making the centre of the remnant brighter in O \textsc{viii}.  Comparing Fig. \ref{fig_oxygen} with Fig. \ref{fig_oxygen_nc}, we see that thermal conduction played a major role in making the SNRs centre bright in O \textsc{viii} at later epochs in dense ISMs.  Our simulations therefore demonstrate the utility of using O \textsc{viii} as a diagnostic of remnant age.  We hasten to point out that our present conclusions for the high-stage ions are limited by our use of equilibrium cooling.  The ionisation fractions of the different oxygen ions could change in the presence of non-equilibrium physics, and a future round of simulations will examine whether this trend is preserved when such physics are included.

\section{Conclusion}\label{section_conclusion}

Anisotropic thermal conduction decreases the average temperature of hot gas by half an order of magnitude over a period of a few million years, and increase the mean density by a similar amount.  The temperature in the hottest portions of the hot gas bubble is reduced by an entire order of magnitude when thermal conduction is included, with a corresponding increase in the density.  These changes affect the emission in high-stage ions rather strongly.  In particular, the distribution of O \textsc{viii} in a denser ISM transitions from shell-bright at early epochs to centre-bright at later epochs when the physics of thermal conduction is included.  

In denser ISMs (such as in proximity to giant molecular clouds, in the bulges of spiral galaxies and in starburst galaxies) the higher density and lower temperatures lead to the development of centre-bright soft x-ray emission at late stages in the evolution of the SNR.  This confirms that it is possible to produce MM SNRs in denser environments through the process of thermal conduction alone.  Our results are concordant with the observations of \citet{rho_petre98}, who find that the majority of MM SNRs are found around dense molecular gas .

We have also shown that the choice of x-ray waveband in which the remnant is imaged can play an important role in deducing whether the remnant is in a mixed-morphology stage or not.  If soft x-rays are used, a remnant expanding into a dense ISM will not show itself as centre-bright until rather late in its evolution.  On the other hand, the same remnant would show a mixed morphology at reasonably early epochs when imaged in hard x-rays.  Remnants expanding into a less dense ISM do not appear centre-bright in either soft or hard x-rays until rather late in their evolution.  Even at those epochs, the hard x-rays would more likely show a centre-bright structure than soft x-rays.

\section*{Acknowledgements}
The authors warmly acknowledge discussions with R. Benjamin.  DSB acknowledges support via NSF grants AST-005569-001 and NSF-PFC grant PHY02-16783.  The simulations were performed at UND and NCSA.  This work has made use of the NASA Astrophysics Data System abstract database.
\bibliographystyle{apj}
\bibliography{ref}

\label{lastpage}

\end{document}